\documentclass[seceq]{ptptex}

\usepackage{graphicx}

\title{Similarity solutions of Fokker-Planck equation with moving boundaries}

\author{C.-L.  \textsc{Ho}}

\inst{
Department of Physics, Tamkang University,
Tamsui 25137, Taiwan, R.O.C\\ 
~\\
May 03, 2012}

\abst{
In this work we present new exact  similarity solutions with moving boundaries of the Fokker-Planck
equation having both time-dependent drift and diffusion coefficients. }

\begin{document}

\maketitle\section{Introduction}

One of the basic tools which
is widely used for studying the effect of fluctuations in
macroscopic systems is the Fokker-Planck equation (FPE) \cite{RIS:1996}.
This equation has found applications in such diverse areas as
physics, chemistry, hydrology, biology, finance and
others. Because of its broad applicability, it is therefore of
great interest to obtain solutions of the FPE for various physical
situations.

Generally, it is not easy to find analytic solutions of the FPE, except in a few simple cases,
such as linear drift and constant diffusion coefficients. In most cases, one can only solve the equation
approximately, or numerically.  Most of these methods, however, are concerned
only with FPEs with time-independent diffusion and drift
coefficients (for a review of these methods, see eg. Ref. ~\citen{RIS:1996}).

Solving the FPEs with time-dependent drift and/or diffusion
coefficient is in general an even more difficult task. It is
therefore not surprising that the number of papers on such kind of
FPE is far less than that on the FPE with time-independent
coefficients.  Some recent works on the FPE with time-dependent
diffusion coefficients appear in
\citen{GMNT:2005,KSF:2005,GNT:2009}, and works involving
time-dependent drift coefficients can be found in
\citen{LM:2000,HY:2008,LH:2011}. Refs.
\citen{WH:1980,OK:1985,SS:1999} consider FPEs with both
time-dependent diffusion and drift coefficients. The symmetry
properties of the one-dimensional FPE with arbitrary coefficients
of drift and diffusion are investigated in \citen{SS:1999}. Such
properties may in some cases allow one to transform the FPE into
one with constant coefficients.

In Ref.~\citen{LH:2011} we have considered, within
the framework of a perturbative approach, the similarity solutions
of a class of FPEs  which have constant diffusion coefficients and
small time-dependent drift coefficients.
The solvability
of the FPE with both time-dependent drift and
diffusion coefficients by means of the similarity method is explored in Ref.~\citen{LH:2012} .
One advantage of the similarity method is that it allows one to
reduce the FPE to an
ordinary differential equation which may be easier to solve, provided that the FPE
possess proper scaling property under certain scaling
transformation of the basic variables.  Some interesting exactly solvable cases of such FPE on the real line $x\in (-\infty,\infty)$ and
the half lines $x\in [0,\infty)$ and $x\in (-\infty,0]$  were considered in Ref.~\citen{LH:2012}.
These domains admit similarity solutions because their boundary points are the fixed points of the scaling transformation considered.
This indicates that similarity solutions are not possible for other finite domains.

It is therefore natural to suspect that similarity solutions of  FPE on a finite domain may be possible, if its boundary points scale accordingly.
We are thus led to consider FPE with moving boundaries. The purpose of this note is to present three new classes of such FPEs.
To the best of our knowledge, FPEs with time-dependent coefficients and moving boundaries have not been discussed before.

\section{Scaling of Fokker-Planck equation}

We first recapitulate the scaling form of the FPE.
The general form of the FPE in $(1+1)$-dimension is
\begin{eqnarray}\label{E2.1}
\frac{\partial W(x,t)}{\partial t}=\Big[-\frac{\partial}{\partial x}
D^{(1)}(x,t)+\frac{\partial^2}{\partial x^2}D^{(2)}(x,t)\Big]W(x,t)\;,
\end{eqnarray}
where $W(x,t)$ is the probability distribution function,
$D^{(1)}(x,t)$ is the drift coefficient and $D^{(2)}(x,t)$ the
diffusion coefficient. The drift coefficient represents the
external force acting on the particle, while the diffusion
coefficient accounts for the effect of fluctuation. $W(x,t)$ as a
probability distribution function should be normalized, $i.e.$,
$\int_{\textstyle\mbox{\small{domain}}}W(x,t)\,dx=1$ for $t\geq
0$.

We shall be interested in seeking similarity solutions of the FPE, which are possible if
the FPE possesses certain scaling symmetry.

Consider the scale transformation
\begin{align}\label{E2.2}
&\bar{x}=\varepsilon^a x\;\;\;,\;\;\;\bar{t}=\varepsilon^b t,
\end{align}
where $\varepsilon$, $a$ and $b$ are real parameters. Suppose
under this transformation, the probability density function and
the two coefficients scale as
\begin{align}\label{E2.3}
\bar{W}(\bar{x},\bar{t})=\varepsilon^c W(x,t),~
\bar{D}^{(1)}(\bar{x},\bar{t})=\varepsilon^d
D^{(1)}(x,t), ~\bar{D}^{(2)}(\bar{x},\bar{t})=\varepsilon^e
D^{(2)}(x,t).
\end{align}
Here $c$, $d$ and $e$ are also some real parameters.  It can be checked that
the transformed equation in terms of the new variables has the same functional form as eq.(\ref{E2.1})
if the scaling indices satisfy $b=a-d=2a-e$. In this case, the second order FPE can  be transformed into an
ordinary differential equation which may be easier to solve. Such reduction is effected
through a new independent variable (called similarity variable), which is
certain combinations of the old independent variables $z$ such that
it is scaling invariant, i.e., no appearance of parameter
$\varepsilon$, as a scaling transformation is performed. Here
the similarity variable $z$ is defined by
\begin{align}\label{E3.1}
z\equiv\frac{x}{t^{\alpha}}\;,\;\;\;\mbox{where}\;\;\;
\alpha=\frac{a}{b}\;\;\;\mbox{and}\;\;\;a\;,b\neq 0\;.
\end{align}
For $a\;,b\neq 0$, one has $\alpha\neq 0\;,\infty$.

The general scaling form of the probability density function
$W(x,t)$ is $W(x,t)=(t^{\delta_1}/x^{\lambda_1})y(z)$, where
$\lambda_1$ and $\delta_1$ are two real parameters and $y(z)$ is a
scale-invariant function under the same transformation
(\ref{E2.2}). From the assumed scaling behavior of $W(x,t)$ in
(\ref{E2.3}), we have $-c/a=\lambda_1-(\delta_1/\alpha)$. Without
loss of generality and for clarity of presentation, we take
the parameters $(\lambda_1,\delta_1)=(0,\alpha c/a)$. This
gives
\begin{align}\label{E3.2}
W(x,t)=t^{\alpha\frac{c}{a}}y(z)\;,
\end{align}
where $y(z)$ is a function of $z$. The normalization of the distribution function is
\begin{align}\label{E3.5}
\int_{\mbox{\small{domain}}}\,W(x,t)\,dx=\int_{\mbox{\small{domain}}}\,
\Big[t^{\alpha(1+\frac{c}{a})}\,y(z)\Big]\,dz=1\;.
\end{align}
For the above relation to hold at all $t\geq 0$, the power of $t$
should vanish, and so one must have $c=-a$, and thus
\begin{equation}
W(x,t)=t^{-\alpha}y(z). \label{W-scale}
\end{equation}
Similar consideration leads to the following scaling forms of the
drift  and diffusion coefficients
\begin{align}\label{E3.3}
D^{(1)}(x,t)=t^{\alpha-1}\rho_1(z)\;\;\;,\;\;\;D^{(2)}(x,t)=t^{2\alpha-1}\rho_2(z)\;,
\end{align}
where $\rho_1(z)$ and $\rho_2(z)$ are scale invariant functions of $z$.

With eqs.~(\ref{E3.1}), (\ref{E3.2}) and (\ref{E3.3}), the FPE is
reduced to
\begin{align}\label{E3.4}
\rho_2(z)\,y''(z)+\Big[2\rho_2'(z)-\rho_1(z)+\alpha z\Big]\,y'(z)
+\Big[\rho_2''(z)-\rho_1'(z)+\alpha \Big]\,y(z)=0\;,
\end{align}
where the prime denotes the derivative with respect to $z$.
It is really interesting to realize  that eq.~(\ref{E3.4}) is exactly integrable.
Integrating it once, we get
\begin{equation}
\rho_2(z) y^\prime(z) + \left[\rho_2^\prime (z) - \rho_1 (z)+
\alpha z\right] y(z)=C, \label{y1}
\end{equation}
where $C$ is an integration constant.
Solution of eq.~(\ref{y1}) is
\begin{eqnarray}
y(z)&=&\left(C^\prime+C\int^z dz \frac{e^{-\int^z\,dz f(z)}}{\rho_2(z)}\right)\,\exp\left(\int^zdz f(z)\right),\nonumber\\
f(z)&\equiv& \frac{\rho_1(z)-\rho_2^\prime(z)-\alpha
z}{\rho_2(z)},~~~\rho_2(z)\neq 0,
\label{y-soln}
\end{eqnarray}
where $C^\prime$ is an integration constant.

To proceed further, let us consider the boundary conditions of
the probability density $W(x,t)$ and the associated probability
current density $J(x,t)$.
From the continuity equation
\begin{equation}
\frac{\partial}{\partial t}W(x,t)=-\frac{\partial}{\partial
x}\,J(x,t)\;,\label{E4.5.2}
\end{equation}
we have
\begin{equation}
J(x,t)=D^{(1)}(x,t)\,W(x,t)-\frac{\partial}{\partial
x}\,\Big[D^{(2)}(x,t)\,W(x,t)\Big]\;.\label{E4.5.3}
\end{equation}
Using eqs.~(\ref{E3.1}), (\ref{W-scale}) and (\ref{E3.3}), we get
\begin{equation}
J(x,t)=t^{-1} \left[\left(\rho_1(z)-\rho_2^\prime(z)\right)y(z)
-\rho_2 (z) y^\prime(z)\right]. \label{J-1}
\end{equation}
From eq.~(\ref{y1}), we can reduce the above equation to
\begin{equation}\label{E4.5.4a}
J(x,t)=\frac{1}{t}\left[\alpha\,z\,y(z)-C\right]=\frac{1}{t}\left[\alpha\,x\,W(x,t)-C\right],
\end{equation}
where $W(x,t)=t^{-\alpha}\,y(z)$ and $z=x/t^{\alpha}$ have been
used in obtaining the second expression.

In this paper we consider boundaries which are impenetrable.  Hence we must have
\begin{eqnarray}
W(x,t)|_{\rm boundary}=J(x,t)|_{\rm boundary}=0.
 \label{Conditions}
\end{eqnarray}
These conditions imply that $C=0$, and that $J(x,t)$ is
proportional to $W(x,t)$ and $x$.

With $C=0$, the probability density function $W(x,t)$ is given by
\begin{eqnarray}
W(x,t)=A t^{-\alpha} \exp\left(\int^z
dz\,f(z)\right)_{z=\frac{x}{t^\alpha}},\label{W-soln}
\end{eqnarray}
where $A$ is the normalization constant.  It is
interesting to see that the similarity solution of the FPE can be
given in such an analytic closed form.  Exact similarity solutions
of the FPE can be obtained as long as $\rho_1(z)$ and $\rho_2(z)$
are such that the function $f(z)$ in eq.~(\ref{W-soln}) is an
integrable function and the resulted $W(x,t)$ is normalizable.
Equivalently, for any integrable function $f(z)$ such that
$W(x,t)$ is normalizable, if one can find a function $\rho_2(z)$
($\rho_1(z)$ is then determined by $f(z)$ and $\rho_2(z)$), then
one obtains an exactly solvable FPE with similarity solution given
by (\ref{W-soln}).  Some interesting cases of such FPE on the real line $x\in (-\infty,\infty)$ and
the half lines $x\in [0,\infty)$ and $x\in (-\infty,0]$  were discussed in Ref.~\citen{LH:2012}.

That the half and whole lines can accommodate similarity solutions which allow $x$ to scale is obvious:
the points $x=0$ and $x=\pm\infty$ are the fixed points of the scale transformation (\ref{E2.2}).
Other finite intervals do not allow similarity solutions, unless their end points scale accordingly. We are thus led to
FPE with moving boundaries.

In what follows we shall present three classes of exactly solvable FPE of such kind.

\section{Class I: two moving boundaries}

We consider a finite domain $x_1(t)\leq x \leq x_2(t)$ with impenetrable moving boundaries at $x_k(t)~(k=1,2)$.
We want the transformed FPE in $z$-space can be exactly solvable. So the simplest choice is such that in the $z$-space the
boundary points of the corresponding domain are static.  This implies $ z_k = x_k(t)/t^\alpha (k=1,2)$ are constants.
We will assume this choice below.  We note here that the fixed  domains admitting similarity solutions considered  in
Ref.~\citen{LH:2012} correspond to the choice $z=x(t)/t^\alpha=0,\pm\infty$, which are just the fixed points of the scaling transformation.

Now let us assume the function $f(z)$ to have the following form in the physical domain in the $z$-space:
\begin{eqnarray}
f(z)=\frac{a_1}{z-z_1}-\frac{a_2}{z_2-z},~~a_1,\,\,a_2>0, ~z_1\leq z\leq z_2.
\end{eqnarray}
This leads to a choice of $\rho_1(z)$ and $\rho_2(z)$:
\begin{align}
\rho_2 (z)&=(z-z_1)(z_2-z),\\
\rho_1 (z)&=(\alpha-a_1-a_2-2)z+(a_1+1)z_2 + (a_2+1)z_1
\end{align}
for $z_1\leq z\leq z_2$, and $\rho_1 (z),\,\rho_2 (z)=0$ otherwise.
The function $y(z)$ in the physical domain is
\begin{eqnarray}
y(z)=A(z-z_1)^{a_1} (z_2-z)^{a_2}.
\end{eqnarray}
Here the normalisation constant $A$ is given by
\begin{eqnarray}
A=[(z_2-z_1)^{a_1+a_2+1}B(a_1+1,a_2+1)]^{-1},
\end{eqnarray}
where $B(x,y)$ is the Beta function.

The probability density function is
\begin{eqnarray}
W(x,t)=
\left\{
\begin{array}{ll}
 \frac{A}{t^\alpha}\left(\frac{x}{t^\alpha}-z_1\right)^{a_1}\left(z_2-\frac{x}{t^\alpha}\right)^{a_2},&z_1t^\alpha\leq x\leq z_2 t^\alpha\\
0,&{\rm otherwise}
 \end{array}\right..
\end{eqnarray}
There are three subclasses:
\begin{align}
({\rm i})~ & z_1,\, z_2>0 ~ (z_1,\,z_2<0);\\
({\rm ii})~ & z_1=0,\, z_2>0 ~ (z_1<0,\, z_2=0);\\
({\rm iii})~ & z_1<0, \, z_2>0.
\end{align}
The situations given in the brackets correspond to mirror images of the corresponding classes with an appropriate change of parameters.
In subclass (ii), $z_1=0$ is a fixed point of the scale transformation, and can be considered as a special case of Case II to be discussed below.

In Figs.~1 to 3 we show figures for  subclass (i) and (iii).   For Figs.~1 and  3 we set $\alpha>0$, while for Fig.~2 we have $\alpha<0$. It is seen that
for $\alpha>0 (<0) $, the boundaries move away from (toward) the origin (except when the boundary is a fixed point).
The same pattern is also exhibited in the other classes.

\begin{figure}
 \centering
 \includegraphics{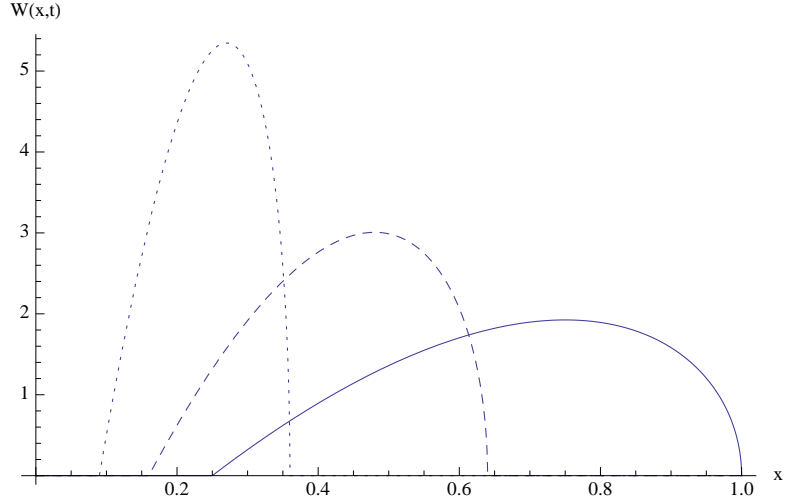}
\caption{$W(x,t)$ vs $x$ for Case-I(i) with $\alpha=2, z_1=1, z_2=4,a_1=1$ and $a_2=1/2$ for $t=0.3$ (dotted line), $0.4$ (dashed line) and $0.5$ (solid line). }
\label{fig.1}
\end{figure}

\begin{figure}
 \centering
 \includegraphics{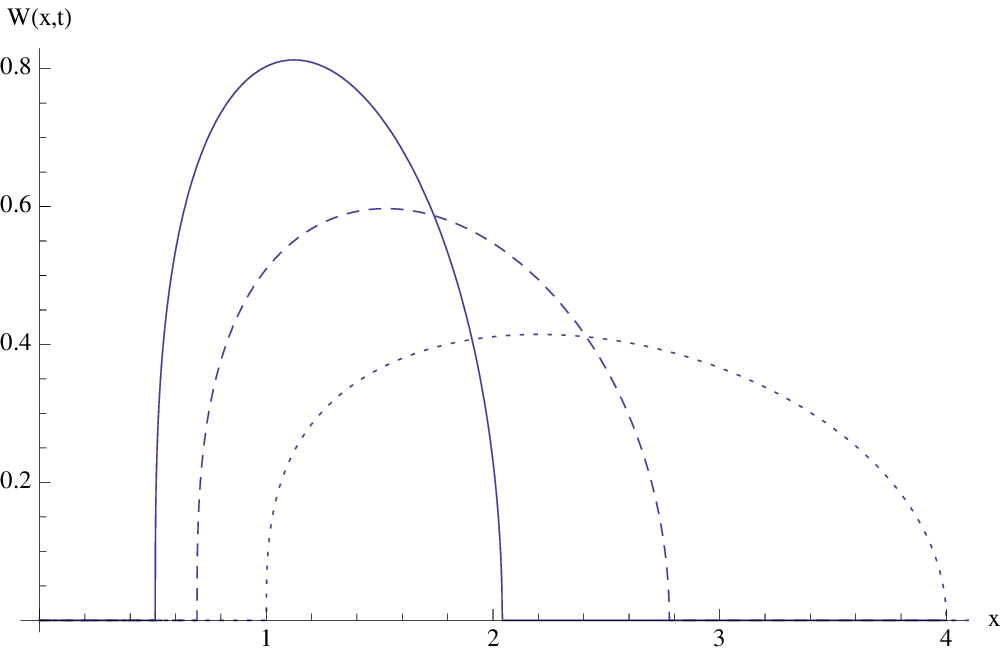}
\caption{$W(x,t)$ vs $x$ for Case-I(i) with $\alpha=-2, z_1=1, z_2=4,a_1=1/3$ and $a_2=1/2$ for $t=1.0$ (dotted line), $1.2$ (dashed line) and $1.4$ (solid line). }
\label{fig.2}
\end{figure}

\begin{figure}
 \centering
 \includegraphics{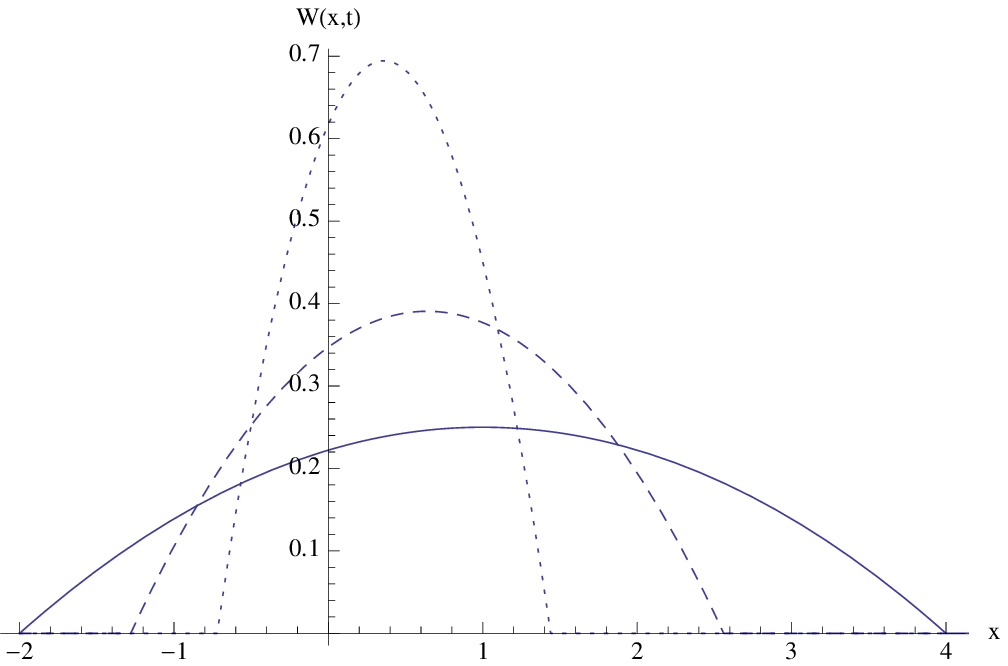}
\caption{$W(x,t)$ vs $x$ for Case-I(iii) with $\alpha=2, z_1=-2, z_2=4$ and $a_1=a_2=1$ for $t=0.6$ (dotted line), $0.8$ (dashed line) and $1.0$ (solid line). }\label{fig.3}
\end{figure}

\section{Class II: One moving boundary with $x=0$ a fixed point}

We now  consider the case on the positive half-line with $x=0$ a fixed point. The case with the moving front in the negative half-line is simply the
mirror images of the situation discussed here  with an appropriate change of parameters.

The function $y(z)$ is given by
\begin{eqnarray}
f(z)=\frac{a_1}{z}-\frac{a_2}{z_2-z}+\beta,~~a_1,\,\,a_2>0, ~0\leq z\leq z_2,~\beta: {\rm real}.
\end{eqnarray}
This leads to a choice of $\rho_1 (z)$ and $\rho_2 (z)$:
\begin{align}
\rho_2  (z)& =z(z_2-z),\\
\rho_1 (z) & =-\beta z^2+(\alpha-a_1-a_2-2+\beta z_2)z+(a_1+1)z_2
\end{align}
for $0\leq z\leq z_2$, and $\rho_1,\,\rho_2=0$ otherwise.
The function  $y(z)$ is
\begin{eqnarray}
y(z)=A z^{a_1} (z_2-z)^{a_2} e^{\beta z},
\end{eqnarray}
where the normalisation constant $A$ is given by
\begin{eqnarray}
A=[z_2^{a_1+a_2+1}B(a_1+1,a_2+1)
\, \,_1F_1(a_1+1;a_1+a_2+2;\beta z_2)]^{-1},
\end{eqnarray}
where $\,_1F_1(\mu;\nu; z)$ is the Kummer confluent hypergeometric function.

The probability density function is
\begin{eqnarray}
W(x,t)=
\left\{
\begin{array}{ll}
 \frac{A}{t^\alpha}\left(\frac{x}{t^\alpha}\right)^{a_1}\left(z_2-\frac{x}{t^\alpha}\right)^{a_2}e^{\beta \frac{x}{t^\alpha}},& 0\leq x\leq z_2 t^\alpha\\
0,&{\rm otherwise}
 \end{array}\right. .
\end{eqnarray}
A representative situation of this class is depicted in Fig.~4.

\begin{figure}
 \centering
 \includegraphics{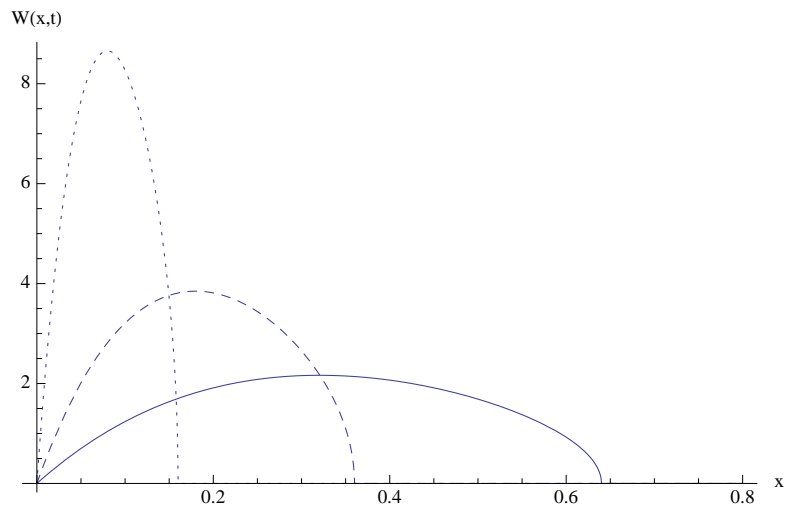}
\caption{$W(x,t)$ vs $x$ for Case-II with $\alpha=2, z_2=1,a_1=1, a_2=1/2$ and $\beta=-1$ for $t=0.4$ (dotted line),
 $0.6$ (dashed line) and $0.8$ (solid line). }\label{fig.4}
\end{figure}

\section{Class III: One moving boundary with $x=\infty$ a fixed point}

This last class represents the case on the positive half-line with $x=\infty$ a fixed point. The situation with  $x=-\infty$ a fixed point is the
mirror image of the situation discussed here  with an appropriate change of parameters.

For this class the function $y(z)$ is taken to be
\begin{eqnarray}
f(z)=\frac{a_1}{z-z_1}-\frac{a_2}{z_2}-\beta,~~a_1,\,\,a_2>0, ~z_1\leq z,~\beta>0.
\end{eqnarray}
The corresponding  choice of $\rho_1 (z)$ and $\rho_2 (z)$ is:
\begin{align}
\rho_2  (z)& =(z-z_1)z,\\
\rho_1  (z)& =-\beta z^2+(\alpha+a_1+a_2+2+\beta z_1)z-(a_2+1)z_1
\end{align}
fo $z_1\leq z$, and $\rho_1,\,\rho_2=0$ otherwise.
The $y(z)$ is
\begin{eqnarray}
y(z)=A (z-z_1)^{a_1} z^{a_2} e^{-\beta z} .
\end{eqnarray}
Here the normalisation constant $A$ is given by
\begin{eqnarray}
A=[\beta^{-\frac{a_1+a_2+2}{2}}z_1^{\frac{a_1+a_2}{2}} \Gamma(a_1+1)e^{-\frac12 \beta z_1}
 \, W_{\frac{a_2-a_1}{2}, -\frac{a_1+a_2+1}{2}} (\beta z_2)]^{-1},
\end{eqnarray}
where $\Gamma(z)$ and $W_{\mu,\nu}( z)$ are the Gamma and Whittaker functions, respectively .

The probability density function is
\begin{eqnarray}
W(x,t)=
\left\{
\begin{array}{ll}
 \frac{A}{t^\alpha}\left(\frac{x}{t^\alpha}-z_1\right)^{a_1}\left(\frac{x}{t^\alpha}\right)^{a_2}e^{-\beta \frac{x}{t^\alpha}},& z_1 t^\alpha \leq x\leq \infty\\
0,&{\rm otherwise}
 \end{array}\right. .
\end{eqnarray}
Fig.~5 shows the evolution of $W(x,t)$ for some choice of parameters.
\begin{figure}
 \centering
 \includegraphics{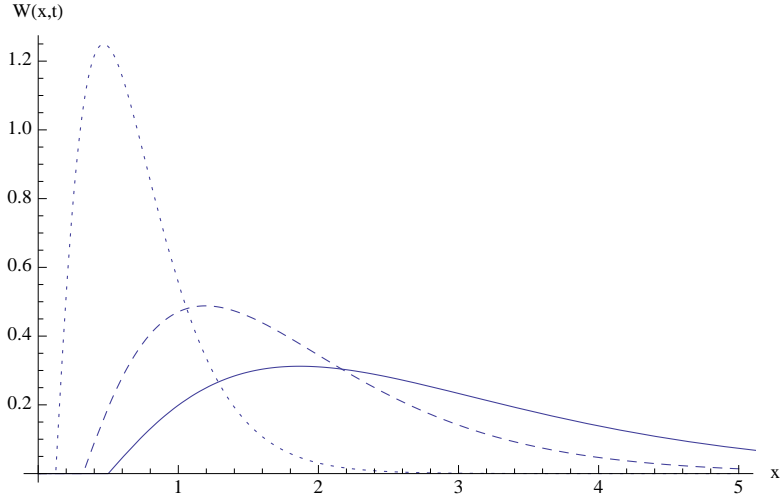}
\caption{$W(x,t)$ vs $x$ for Case-III with $\alpha=2, z_1=0.5, a_1=1, a_2=1/2$ and $\beta=1$ for $t=0.5$ (dotted line), $0.8$ (dashed line) and $1.0$ (solid line). }\label{fig.5}
\end{figure}

In summary, we have found that similarity solutions of the FPE are possible with moving boundaries, provided that the boundaries points scale appropriately.
Three new classes of exactly solvable FPE's with moving boundaries are presented.

\section*{Acknowledgments}

This work is supported in part by the
National Science Council (NSC) of the Republic of China under
Grant NSC-99-2112-M-032-002-MY3.


\end{document}